\documentclass[aps,showpacs,twocolumn,superscriptaddress,floatfix]{revtex4}
\usepackage{graphicx}
\usepackage{epsfig}
\usepackage{amsfonts}
\usepackage{amsmath,amssymb}
\usepackage{bm}
\usepackage{booktabs}

\renewcommand{\vec}[1]{\bm{#1}}

\newcommand{\disregard}[1]{}

\begin{document}

\title{Isospin-symmetry restoration within the nuclear density functional theory: formalism and applications}

\author{W. Satu{\l}a}
\affiliation{Institute of Theoretical Physics, University of Warsaw, ul. Ho\.za
69, 00-681 Warsaw, Poland}

\author{J. Dobaczewski}
\affiliation{Institute of Theoretical Physics, University of Warsaw, ul. Ho\.za
69, 00-681 Warsaw, Poland}
\affiliation{Department of Physics, P.O. Box 35 (YFL),
FI-40014 University of Jyv\"askyl\"a, Finland}

 \author{W. Nazarewicz}
\affiliation{Department of Physics \&
  Astronomy, University of Tennessee, Knoxville, Tennessee 37996}
\affiliation{Physics Division, Oak Ridge National Laboratory, P.O. Box
  2008, Oak Ridge, Tennessee 37831}
\affiliation{Institute of Theoretical Physics, University of Warsaw, ul. Ho\.za
69, 00-681 Warsaw, Poland}

 \author{M. Rafalski}
\affiliation{Institute of Theoretical Physics, University of Warsaw, ul. Ho\.za
69, 00-681 Warsaw, Poland}

\date{\today}

\begin{abstract}
Isospin symmetry of atomic nuclei is explicitly broken by the charge-dependent interactions, primarily the Coulomb force. Within the nuclear density functional theory, isospin is also broken spontaneously. We propose a projection scheme rooted in a mean field theory,  that allows the consistent treatment of isospin breaking in both ground and exited nuclear states. We demonstrate that
this scheme is essentially free from spurious divergences plaguing particle-number and angular-momentum restoration approaches.
Applications of the new technique include
excited high-spin states in medium-mass $N$=$Z$ nuclei, such as superdeformed bands and  many-particle-many-hole terminating states.
\end{abstract}

\maketitle

\section{Introduction}\label{intro}

Density functional theory (DFT) in its original formulation \cite{[Hoh64w],[Koh65w]} and its extended versions
\cite{[Lev79],[Lev82],[Lie83],[Lie85]} has
became a universal approach
to compute the ground-state (g.s.) and excited \cite{[Eng88a],[Gor96a]} configurations of many-electron systems held
together by an external one-body
potential in condensed-matter, atomic, and molecular physics.
The DFT strategy has also been  used in the area of nuclear
structure. The nuclear DFT, a natural extension of the self-consistent mean-field (MF) theory \cite{[Ben03],[Lal04]} is a tool of choice for computations of g.s. properties
and low-lying excitations of medium-mass and heavy nuclei.

There are a number of differences between the electronic DFT and nuclear DFT. First of all, because of the absence of external potential, nuclei are self-bound systems, and this creates conceptual problems due to the distinction between intrinsic (symmetry-broken)  and laboratory-system densities. This problem can be treated within the extended DFT framework of Levy-Lieb \cite{[Lev79],[Lev82],[Lie83],[Lie85],[Pro05]}, see Refs.~\cite{[Eng07],[Gir08],[Gir08a],[Mes09a]}.
The second difference is related to the effective interaction.
A complicated form of the effective nucleon-nucleon
interaction, as well as complexity of the nuclear matter
saturation-mechanism (pertaining to three-body forces),
give rise to the presence of higher-order terms in the
nuclear energy density functional (EDF). Hence, the EDF cannot be constructed
in a model-independent {\it ab initio\/} way based solely on
bulk properties of infinite-medium augmented by
exchange and gradient corrections, as it is usually done for electronic systems.
The current best nuclear EDFs are still phenomenological in nature,  either inspired
by a MF averaging of the effective
interaction such as  the zero-range Skyrme~\cite{[Sky56xw]} force,
or  based on systematic expansion involving symmetry-constrained gradient
terms~\cite{[Car08]}. By construction, the nuclear EDF is characterized
by a certain number of free parameters which are fitted directly
to empirical data.

The third major difference between electronic and nucleonic DFT, addressed in
this study,  is the presence of  two types of fermions
and  the charge independence of
the nuclear interactions giving rise to the isospin symmetry.
Since the isospin symmetry is violated essentially by
the Coulomb interaction, which is much weaker than the strong
interaction between nucleons, many effects associated with the isospin
breaking in nuclei can be treated in a perturbative way, making
the formalism of isotopic spin a very powerful concept in nuclear
structure and reactions \cite{[Wil69],[War06a]}.

Up to electromagnetic effects,  the isospin symmetry should be conserved by
elementary excitations of  nuclei. This is not the case for the elementary  excitations modes of the nuclear Hartree-Fock
(HF) or Kohn-Sham theory, i.e.,  proton and/or neutron particle-hole (p-h) excitations. The independent-particle wave function
manifestly breaks
the isospin symmetry in the g.s.\
configurations of odd-odd $N$=$Z$ nuclei and in all other but isoscalar
excited states.
Two prominent examples
are discussed in this work: superdeformed (SD)
bands  in a doubly magic nucleus
$^{56}$Ni~\cite{[Rud99fw]} and  terminating states in $N$=$Z$
$A\sim$40 nuclei~\cite{[Zal07]}.

The paper is organized as follows.
We begin in Sec.~\ref{oddodd} with general
discussion of g.s.\ configurations of odd-odd
$N$=$Z$ nuclei and p-h excitations in even-even (e-e) nuclei,
so as to qualitatively introduce problems faced by  MF-based
theories around  $N$=$Z$  and to
provide motivation for the necessity of isospin restoration.
Section~\ref{theory} contains
a detailed presentation of the isospin-projected DFT approach. In particular, we
demonstrate that this approach
is essentially free from problems related to uncompensated poles plaguing
projection techniques \cite{[Dob07d]}; hence,
it requires no further regularization \cite{[Lac09]}.
Applications of the isospin-projected DFT approach to the
SD bands in $^{56}$Ni and terminating states in $N$=$Z$ isotopes of Ca, Sc, Ti, V,
and Cr, are presented in  Secs.~\ref{ni56} and~\ref{term}, respectively.
Finally,  Sec.~\ref{summ} contains the conclusions of this work.

\section{Isospin symmetry violation in a mean-field approach}\label{oddodd}

\begin{figure}
\includegraphics[width=0.8\columnwidth, clip]{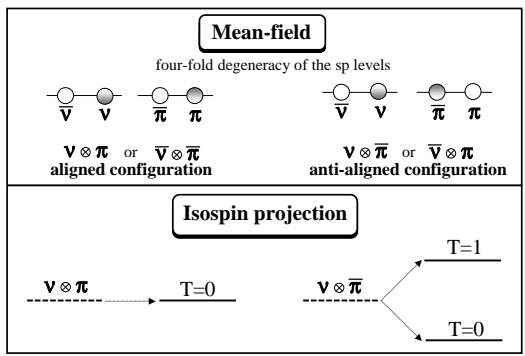}
\caption{The upper panel schematically shows two possible g.s.\ configurations of an odd-odd $N$=$Z$ nucleus,
as described by the conventional deformed MF theory. These
 degenerate configurations are called aligned (left) and anti-aligned (right),
depending on what levels are  occupied by the valence particles. The lower panel shows what happens if
the isospin-symmetry is restored. The aligned configuration is isoscalar; hence,
it is insensitive to the
isospin projection. The anti-aligned configuration
represents a mixture of the $T$=0 and $T$=1 states. The isospin projection removes the degeneracy by
shifting the $T$=0 component down.
}\label{fig1}
\end{figure}

Let us begin  by exposing serious problems with the MF
description of g.s.\ configurations of odd-odd (o-o) $N$=$Z$ nuclei.
If, for the sake of simplicity, the Coulomb and time-odd
polarization effects are disregarded and proton-neutron symmetry is conserved, a deformed MF approach naturally leads to four-fold
degenerate (isospin and Kramers) single-particle (s.p.) levels. Consequently, the MF
g.s.\ configuration of the o-o $N$=$Z$ nucleus is not uniquely defined and
depends on  occupation of specific levels.
As shown in the upper panel of Fig.~\ref{fig1},
the valence proton and neutron can be arranged in two
distinctively different ways. Indeed,  configurations of the type
shown on the left-hand side (aligned configurations), are
symmetric in spin-space coordinates and, therefore, anti-symmetric
(isoscalar) in  isospin coordinates. The  anti-aligned configurations
depicted on the right-hand side, have a mixed
symmetry in spin-space coordinates and, therefore,  also
in isospin coordinates.

Because of their isoscalar character,  aligned configurations are not affected by the isospin projection. On the other hand, the projection lifts
the degeneracy  of $T$=0 and $T$=1 components of aligned configurations  as it is illustrated in the lower panel of Fig.~\ref{fig1}.
Due to the repulsive character of the nuclear symmetry energy,
the isovector (isoscalar) components of the anti-aligned configurations
are shifted up (down) in energy. Hence, isospin restoration  changes the structure of the ground states of o-o $N$=$Z$ nuclei
without affecting the ground states of e-e $N$=$Z$ nuclei. In other
words, it does affect the binding-energy staggering along the $N$=$Z$ line.

\begin{figure}[tbp!]
\includegraphics[width=0.80\columnwidth, clip]{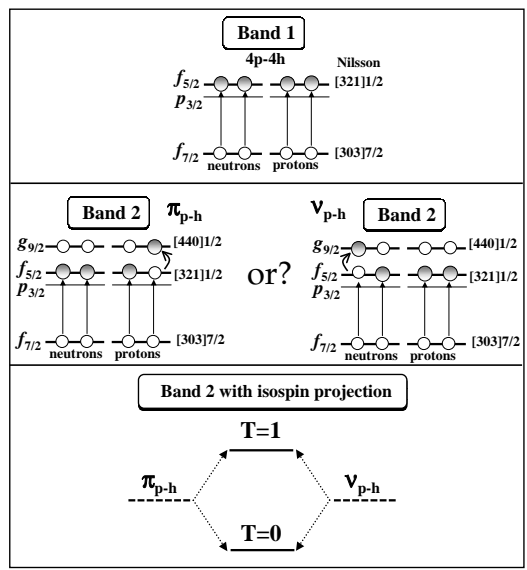}
\caption{The upper panel schematically shows the s.p.\ configuration
corresponding to the SD Band 1 in $^{56}$Ni, which is based on a 4p4h
excitation with respect to the doubly magic core. The middle panel
depicts two possible configurations of Band 2 involving a single
proton (left) or neutron (right) promoted to the $0g_{9/2}$
shell. To facilitate the discussion, we also show the relevant asymptotic
Nilsson quantum numbers associated with  deformed orbitals involved. The lower panel illustrates the isospin splitting
induced by the isospin-symmetry restoration.}\label{fig2}
\end{figure}

Another example that nicely illuminates  problems with isospin encountered in
MF approaches is the case of
two SD bands observed in the doubly magic
nucleus $^{56}$Ni~\cite{[Rud99fw]}. Following Ref.~\cite{[Rud99fw]}, we label them as Band 1 and Band 2.
Band 1 is interpreted as a four-particle
four-hole configuration formed by promoting two protons and two neutrons
from the $0f_{7/2}$ shell to $0f_{5/2}$ shell. Within deformed
MF model, where it is more natural to use the notion of Nilsson orbitals, Band 1
is obtained by emptying the neutron and proton [303]7/2 Nilsson
extruder orbitals and occupying the prolate-driving [321]1/2 levels.
This interpretation, not involving the $0g_{9/2}$ shell, is strongly supported
by both
state-of-the-art shell-model (SM)
calculations~\cite{[Ots98w],[Rud99fw]} in the $0\hbar\omega$ $fp$ space  and the self-consistent cranked-HF
theory~\cite{[Rud99fw]}. Both approaches appear to
reproduce satisfactorily the excitation energy and moment of inertia (MoI)
of this band. The p-h configuration of this band is depicted schematically in the upper
panel of Fig.~\ref{fig2}.

Band 2, on the other hand, cannot be reproduced
within the $fp$ SM space. This band has also a
larger MoI as compared to the Band 1, and it is becoming
yrast around spin $I\sim 12\hbar$.
These two facts
strongly suggest that its structure must involve at least one particle
in the prolate-driving [440]1/2
Nilsson orbital that originates from the $0g_{9/2}$ shell.
According to the calculations of Ref.~\cite{[Rud99fw]}, Band 2 involves
one proton in $\pi$[440]1/2 state. Its configuration, which can be regarded as
one proton p-h excitation with respect to
the reference Band 1, is schematically shown in  Fig.~\ref{fig2} (middle panel, left-hand side). This scenario has been supported by
 the full $pf-$ and $pfg_{9/2}$-SM calculations \cite{[Miz00]}.

The conventional MF interpretation of Band 2  is not fully supported by experiment. Indeed, the MF theory predicts existence of the second band that
is built upon the single neutron occupying the $\nu$[440]1/2 orbital
(middle panel  of Fig.~\ref{fig2}, right-hand side).
This neutron band is predicted to be placed only slightly higher in
energy than the proton band, and this is due to a tiny difference in
the Coulomb energies between these two configurations with,
otherwise, very similar properties. Therefore, it is extremely difficult to understand within MF theory why only one of
these two bands is observed experimentally. Moreover, HF calculations
predict that the $\pi$[440]1/2 band is too high in excitation energy
and that it crosses Band 1 above spin $I\sim 16\hbar$, that is, well
above the empirical crossing. In these discussion, energetic
arguments essentially exclude configurations involving two or more
particles in the $0g_{9/2}$ shell.

Here again the problems with the MF approach can be traced back to
the isospin-symmetry violation by the p-h excitations. Indeed, the MF approach
treats proton $|\pi_{p-h}\rangle$ and neutron $|\nu_{p-h}\rangle$ p-h
excitations as independent elementary excitations, and this manifestly
breaks the isospin symmetry in $N$=$Z$ nuclei.  To make the
elementary excitation modes in $N$=$Z$ nuclei consistent with the
 symmetry,  one needs to symmetrize or
anti-symmetrize the Slater determinants corresponding to the mirror
p-h excitations  $|T=0(1)\rangle \approx \frac{1}{\sqrt{2}}
(|\pi_{p-h}\rangle \pm |\nu_{p-h}\rangle )$. Such wave functions go beyond
the usual MF picture. Restoration of isospin  lifts the
MF degeneracy between  proton and neutron excitations and shifts
the isoscalar (isovector) configurations down (up) in energy, as
illustrated in the lower panel of Fig.~\ref{fig2}. Of course, it is
irrelevant whether the $T$=0 and $T$=1 components are projected from
the proton or neutron state; the
results should be identical up to the tiny polarization effects.

At present, it is not at all clear how to include the isospin correlations
directly into the functional. The most natural method of taking them
it into account is the isospin projected DFT
approach~\cite{[Cau80],[Cau82],[Raf09d],[Sat09a]}, which is also used
in this work. We note that such an approach is
within the provisos of the DFT, whereby the isospin-projected energy
is still a functional of the isospin-unprojected density,
cf.~Refs.~\cite{[She00a],[She02],[Sto07]}.

We finish the discussion in this Section by noticing that even the
isospin-projected DFT approach cannot fully account for the structure of
the g.s.\ of an o-o $N$=$Z$ nucleus. Indeed, within the conventional DFT approach,
the aligned and anti-aligned configurations differ due to
different time-odd (TO) polarization effects in these two configurations.
This polarization  affects the
position of the $T=1,T_z=0$ states
with respect to the $T=1,T_z=\pm 1$ states. While the former states are shifted due to TO polarization, the latter ones --
the $I=0$ ground states of e-e $N-Z=\pm 2$ isotopes -- are not influenced by TO effects  due to time-reversal symmetry conservation.
Hence, the TO terms introduce a very specific source of
the isospin symmetry violation in the intrinsic system.

Moreover, as shown in the lower part of Fig.~\ref{fig1}, the
isospin projected DFT approach always yields
ground states with $T$=0 in o-o $N$=$Z$ nuclei. This is at variance with empirical
data; indeed, it is well known
(see, e.g., Refs.~\cite{[Mar99a],[Vog00],[Jan05a]}) that, with two
exceptions, the isospin of the g.s.\ changes from $T$=0 in light ($A
< 40$) o-o $N$=$Z$ nuclei to $T$=1 in  heavier  ($A > 40$) o-o $N$=$Z$
nuclei. To be able to address this question in the nuclear DFT,
one has to consider simultaneous isospin and angular-momentum
projections and also consider residual proton-neutron interactions between valence particles (either perturbatively or by breaking the proton-neutron symmetry).
This subject will be considered in  forthcoming studies. The focus of this paper is  on high-spin superdeformed and
terminating states, which -- due to large spin polarization and high seniority --
are expected to be less affected by the effects mentioned above.

\section{Theory}\label{theory}

\subsection{Isospin  projection formalism}

Within the MF approximation, the isospin symmetry breaking has two
sources. The spontaneous breaking  of isospin associated with the MF
approximation itself~\cite{[Eng70],[Cau80],[Sat09a]}
is, in this theory, intertwined with the explicit symmetry breaking
due to  the Coulomb interaction. The essence of our
method~\cite{[Raf09d],[Sat09a]} is to retain
the explicit isospin mixing. It is achieved by rediagonalizing the
total effective nuclear Hamiltonian $\hat H_{NN}$ in good-isospin
basis. The basis is generated by using the
projection-after-variation technique, that is, by acting with
the standard one-dimensional isospin-projection operator on
the MF product state
$|\Phi\rangle$:
\begin{eqnarray}\label{eqn:imk}
|TT_z\rangle & = &  \frac{1}{\sqrt{N_{TT_z}}}\hat{P}^T_{T_z T_z} |\Phi\rangle   \nonumber \\
& = & \frac{2T+1}{2 \sqrt{N_{TT_z}}}\int_0^\pi d\beta\sin\beta \; d^{T}_{T_z T_z}(\beta)\;
\hat{R}(\beta)|\Phi\rangle,
\end{eqnarray}
where  $\beta$ denotes the Euler angle
associated with the rotation operator $\hat{R}(\beta)= e^{-i\beta \hat{T}_y}$
about the $y$-axis in the isospace, $d^{T}_{T_z T_z}(\beta)$
is the Wigner function~\cite{[Var88]}, and $T_z =(N-Z)/2$ is the third
component of the total isospin $T$.
The wave function $|\Phi\rangle$ is obtained self-consistently by solving the Skyrme-Hartree-Fock (SHF)
equations.
In the HF limit, the $z$-component $T_z$ is strictly conserved.
The overlap $N_{T T_z}$, or interchangeably the normalization factor
$b_{T,T_z}$, are given by
\begin{eqnarray}
\label{eqn:ovr}
N_{T T_z}  & \equiv &  b_{T,T_z}^2 = \langle \Phi | \hat{P}^T_{T_z T_z} | \Phi \rangle
\nonumber \\
& = & \frac{2T+1}{2}\int_0^\pi d\beta \sin\beta \; d^{T}_{T_z T_z}
(\beta ) \; {\mathcal N}(\beta),
\end{eqnarray}
where
\begin{equation}
{\mathcal N}(\beta)  = \langle \Phi| \hat{R}(\beta)| \Phi\rangle
\label{N-ker}
\end{equation}
stands for the overlap kernel. The good-isospin basis
is spanned by the states $|T,T_z\rangle$ that have tangible
contributions to the MF state
\begin{equation}\label{mix}
|\Phi \rangle = \sum_{T\geq |T_z|} b_{T,T_z} |T,T_z\rangle ,
\quad \sum_{T\geq |T_z|} |b_{T,T_z}|^2 = 1.
\end{equation}
In practice, we retain states having $|b_{T,T_z}|^2>10^{-10}$, which sets the limit
of $T\leq |T_z|+5$.

The total nuclear Hamiltonian consists of the kinetic
energy $\hat T$, the Skyrme interaction $\hat V^{S}$ and the
Coulomb interaction $\hat V^{C}$ that breaks isospin:
\begin{eqnarray}\label{ham}
\hat H_{NN} & = &\hat T + \hat V^{S} + \hat V^{C} \nonumber \\
& \equiv & \hat T + \hat V^{S} +\hat V_{00}^C +  \hat V_{10}^C + \hat V_{20}^C ,
\end{eqnarray}
where
\begin{eqnarray}
     \hat  V_{00}^C(r_{ij}) & = & \; \; \; \frac{1}{4}\frac{e^2}{r_{ij}}
       \left( 1 + \frac{1}{3} \hat {\tau}^{(i)} \circ
                  \hat {\tau}^{(j)} \right) \label{C00} \\
      \hat V_{10}^C(r_{ij}) & = &  - \frac{1}{4}\frac{e^2}{r_{ij}}
          \left( \hat \tau_{10}^{(i)}  +
                  \hat \tau_{10}^{(j)} \right)  \label{C10}   \\
     \hat  V_{20}^C(r_{ij}) & = &  \; \; \; \frac{1}{4}\frac{e^2}{r_{ij}} \;
              \left( \hat \tau_{10}^{(i)}
         \hat \tau_{10}^{(j)} - \frac{1}{3} \hat {{\tau}}^{(i)} \circ
                                            \hat {{\tau}}^{(j)} \right)
                                \label{C20}
\end{eqnarray}
are,  respectively, isoscalar, covariant rank-1 (isovector),
and covariant rank-2 axial (isotensor) spherical
tensor components of the Coulomb interaction. They are constructed by coupling
spherical components of the one-body isospin operator:
\begin{equation}
      \hat  \tau_{10} = \hat \tau_{z}, \quad  \hat \tau_{1\pm 1}
           = \mp \frac{1}{\sqrt{2}}
        \left( \hat \tau_{x} \pm i \hat \tau_{y} \right) ,
\end{equation}
where $\hat \tau_{i},\,i=x,y,z$ denote Pauli matrices and
symbol $\circ$ stands for the scalar product of
isovectors.

The Coulomb interaction is the only source of
the isospin symmetry violation in our model. Charge symmetry breaking
components of the strong interaction and the isovector kinetic
energy (which is quenched as compared to its isoscalar counterpart
by a factor $\frac{\Delta M}{M} \sim 0.001$ due to very small
mass difference $\Delta M$ between the neutron and the proton relative to
mean nucleonic mass $M$) are not taken into account.
Hence, the isoscalar part of the total
Hamiltonian reads $\hat H_{00} = \hat T + \hat V^{S}
+\hat V_{00}^C$. Its  matrix elements can be  cast into
one-dimensional integrals:
\begin{eqnarray}\label{eqn:ker}
 \langle T T_z | \hat H_{00} | T' T_z\rangle
 & = &  \frac{\delta_{TT'}}{N_{TT_z}} \langle \Phi | \hat H_{00}
 \hat{P}^T_{T_z T_z} | \Phi \rangle    \nonumber
\\
 =  \frac{\delta_{TT'}}{N_{TT_z}} \frac{2T+1}{2}\int_0^\pi & d\beta & \sin\beta \; d^{T}_{T_z T_z}(\beta )
\; {\mathcal H}_{00}(\beta)  ,
\end{eqnarray}
involving the Hamiltonian  kernel
\begin{equation}\label{H-ker}
{\mathcal H_{00}}(\beta) = \langle \Phi|\hat{H}_{00} \hat{R}(\beta)|
\Phi\rangle.
\end{equation}

Calculation of matrix elements of the isovector (\ref{C10}) and isotensor (\ref{C20})
components of the Coulomb interaction is slightly more complicated. It appears, however,
that these matrix elements can also be reduced to one-dimensional integrals
over the Euler angle $\beta$:
\begin{equation}\label{coul_me}
\langle T' T_z | \hat V^C_{\lambda 0} |  T T_z \rangle =
\frac{1}{\sqrt{N_{T' T_z} N_{T T_z}}}
\sum_{\mu = -\lambda}^{\mu = \lambda} I_{\lambda \mu} ,
\end{equation}
where
\begin{eqnarray}\label{calki}
  I_{\lambda \mu} & = & (-1)^\mu C^{T' T_z}_{T T_z \lambda 0}
                           C^{T' T_z}_{T T_z -\mu \lambda \mu}    \\
    \times \frac{2T +1}{2}\int_0^\pi & d\beta & \sin\beta \; d^{T}_{T_z-\mu T_z}(\beta)\;
     \langle\Phi | \hat V^C_{\lambda \mu } \hat{R}(\beta)|\Phi\rangle ,  \nonumber
\end{eqnarray}
and symbols $C^{T K}_{T M \lambda \mu}$ stand for the Clebsch-Gordan
coefficients~\cite{[Var88]}. The non-axial ($\mu$$\ne$0)  isovector and isotensor components of
the Coulomb interaction appearing in Eq.~(\ref{calki}) are:
\begin{eqnarray}
    \hat V^C_{1\pm 1}(r_{ij}) & = & - \frac{1}{4}\frac{e^2}{r_{ij}}
      \left( \hat \tau_{1\pm 1}^{(i)} + \hat \tau_{1\pm 1}^{(j)} \right), \\
    \hat V^C_{2\pm 1}(r_{ij}) & = & \; \; \; \frac{1}{4}\frac{e^2}{r_{ij}}
         \frac{1}{\sqrt3} \left( \hat \tau_{1\pm 1}^{(i)} \hat \tau_{1 0}^{(j)}
                    + \hat \tau_{1 0}^{(i)} \hat \tau_{1 \pm 1}^{(j)} \right) , \\
   \hat V^C_{2\pm 2}(r_{ij}) & = & \; \; \; \frac{1}{4} \frac{e^2}{r_{ij}}
   \; \sqrt{\frac{2}{3}} \hat \tau_{1\pm 1}^{(i)}
    \hat \tau_{1\pm 1}^{(j)}.
\end{eqnarray}
Equations (\ref{coul_me})--(\ref{calki}) can be derived by using the
standard transformation rules for a  covariant spherical-tensor
operator $\hat T_{\lambda \mu}$ of rank-$\lambda$ under
the three-dimensional  rotation by the Euler angles
$\Omega = (\alpha, \beta, \gamma)$, that is,
\begin{equation}\label{rotat}
   \hat R(\Omega) \hat  T_{\lambda \mu}  \hat R(\Omega)^\dagger
   = \sum_{\mu'} D^\lambda_{\mu'  \mu} (\Omega) \hat  T_{\lambda \mu'},
\end{equation}
where the SO(3) projection operator  $\hat P^{T}_{K M}(\Omega )$
commutes with spherical tensors as~\cite{[Rod02bw]}
\begin{eqnarray}
  && \hat P^{T_f}_{K_f M_f} \hat  T_{\lambda \mu} \hat P^{T_i}_{M_i K_i}
  = \strut
     \nonumber \\
  &=& C^{T_f M_f}_{T_i M_i \lambda \mu}
   \sum_{M \mu'} (-1)^{(\mu'-\mu)}  C^{T_f K_f}_{T_i M \lambda \mu'}
   \hat  T_{\lambda \mu'} \hat P^{T_i}_{M K_i}.
\end{eqnarray}

The cornerstone of the isospin projection scheme described above is a
calculation of the Hamiltonian and norm kernels and subsequent
one-dimensional
integration over the Euler angle $\beta$. The integrals are calculated
numerically using the Gauss-Legandre quadrature, which is very well suited for this problem provided that the calculated kernels
are non-singular.

\subsection{Inverse of the overlap matrix}

A prerequisite for the calculation of the integrals in Eqs.~(\ref{eqn:ovr}),
(\ref{eqn:ker}), and (\ref{calki})
is the isospin rotation of the Slater determinant.
To this end, one has to perform an independent rotation of
 s.p.\ neutron,
\begin{equation}
   \varphi_i  = \left( \begin{array}{c} \varphi_i^{(n)}  \\
                               0  \end{array}
                                    \right),
          \quad   i=1,2,\ldots , N,
\end{equation}
and proton,
\begin{equation}
   \varphi_i  = \left( \begin{array}{c} 0  \\
                                \varphi_i^{(p)} \end{array}
                                    \right),
          \quad  i=N+1,\ldots , A,
\end{equation}
HF wave functions. The use of the two-dimensional (spinor) representation of isospin constitutes
a natural formalism
for the isospin projection technique.
In this representation, the upper (neutron, $q=1$)
and lower (proton, $q=-1$) components of the isospin-rotated states are given by:
\begin{equation}
\hat{R}(\beta)\; \varphi_i \equiv
\tilde{\varphi_i}(\beta),
\end{equation}
where
\begin{eqnarray}
\tilde{\varphi_i}(\beta) &=&
\left( \begin{array}{r}  \varphi_i^{(n)} \cos\frac{\beta}{2}  \\
                         \varphi_i^{(n)} \sin\frac{\beta}{2} \end{array}
                          \right),
          \quad   i=1,2,\ldots , N, \\
\tilde{\varphi_i}(\beta) &=&
\left( \begin{array}{r} -\varphi_i^{(p)} \sin\frac{\beta}{2}  \\
                         \varphi_i^{(p)} \cos\frac{\beta}{2} \end{array}
                          \right),
          \quad  i=N+1,\ldots , A.
\end{eqnarray}

To calculate the kernel $\langle \Phi| \hat{{\mathcal Q}} | \tilde{\Phi} \rangle$ of an arbitrary
operator $\hat{{\mathcal Q}}$ between two non-orthogonal Slater determinants,
we apply the generalized Wick's theorem,
see, e.g., Ref.~\cite{[Bla86]}. In particular, the norm kernel
can be written in a compact form:
\begin{equation}\label{eqn:non-ovr}
\langle \Phi | \tilde{\Phi}\rangle = \mbox{Det}\; {\tilde {O}},
\end{equation}
where  the overlap matrix is:
\begin{equation}\label{eqn:ovr-ij}
{\tilde O}(\beta)_{ij} =  \int d{\boldsymbol r} \sum_{\sigma , q}
{\varphi_i^*}({\boldsymbol r}, \sigma, q)
{\tilde{\varphi}_j} ({\boldsymbol r},\sigma, q, \beta)
. \end{equation}

To calculate kernels appearing in the projection formalism, one needs to
invert the
overlap matrix  ${\tilde {O}}^{-1}$. This can cause serious problems due to the presence of   singularities~\cite{[Dob07d],[Zdu07]}.
The regularization of kernel singularities is a difficult problem \cite{[Lac09]}. Thus far, a
regularization scheme has been worked out only for a very specific class
of functionals (or effective density-dependent
interactions) solely involving integer powers of local densities \cite{[Lac09]}.
Unfortunately, almost all commonly used  Skyrme and Gogny parameterizations, except for SIII~\cite{[Bei75]},
involve fractional powers of the density. The appearance of singularities prevents us
from using  the local Slater approximation for the
Coulomb exchange. In the present work we treat it  exactly using the method of the Gaussian
decomposition of the Coulomb interaction, as described in
Refs.~\cite{[Dob96],[Dob09d]}.

Compared to the particle number
or the angular momentum projection schemes, isospin projection is a relatively simple procedure. In particular,
the dependence of the inverse matrix
${\tilde {O}}^{-1}(\beta)$ on the isorotation angle $\beta$ can be determined
analytically and this enables us to demonstrate that
the isospin projection is free from kernel singularities. To this end, we write
 the overlap matrix (\ref{eqn:ovr-ij})
 in the form:
\begin{equation}\label{eqn:ovr-ij2}
{\tilde {O}}(\beta) = \left(
\begin{array}{cc}
  \cos\frac{\beta}{2} {I}_N &  -\sin\frac{\beta}{2} {O}  \\
 \sin\frac{\beta}{2} {O}^\dagger &  \cos\frac{\beta}{2} {I}_Z
      \end{array} \right),
\end{equation}
where ${I}_N$ (${I}_Z$) stands for the $N\times N$ ($Z\times Z$)
unit matrix, while
${O} = \langle {\varphi}^{(n)} | {\varphi}^{(p)}
\rangle$ is the rectangular $N\times Z$ overlap matrix of the
neutron and proton s.p.
wave functions. Using the singular value
decomposition (SVD) technique,  the  matrix ${O}$ can be written as:
\begin{equation}\label{SVD}
{O}   =      {W} {D}{V}^\dagger ,
\end{equation}
where ${W}$ is an $N\times Z$  rectangular matrix having orthogonal columns
(${W}^\dagger {W} = {I}_Z$),
${D}$ is a diagonal $Z\times Z$ real and non-negative matrix
(${D}^* = {D}$), and  ${V}$ is a
$Z\times Z$ quadratic unitary matrix
(${V}^\dagger {V} =
{V} {V}^\dagger = {I}_Z$). The SVD
decomposition (\ref{SVD}) further implies that:
\begin{subequations}
\begin{eqnarray}
{O} {O}^\dagger & = &
    {W} {D}^2 {W}^\dagger,
      \\
{O}^\dagger {O} & = &
    {V} {D}^2 {V}^\dagger .
\end{eqnarray}
\end{subequations}
where  ${O} {O}^\dagger$ and ${O}^\dagger {O}$ are matrices of
dimension  $N\times N $ and $Z\times Z$, respectively.

The SVD decomposition allows us to analytically diagonalize the overlap matrix
${\tilde {O}}(\beta)$. Without  loss of generality we assume
that $N\geq Z$. Hence,  the product
${\tilde {O}}(-\beta) {\tilde {O}}(\beta)$ can be written as:
\begin{widetext}
\begin{equation}\label{oo-matrix}
{\tilde {O}}(-\beta) {\tilde {O}}(\beta)
= \left( \begin{array}{cc} \tilde{{W}} & 0  \\
                              0   &  {V} \end{array} \right)
  \left( \begin{array}{cc}  \cos^2 \frac{\beta}{2}{I}_N + \sin^2 \frac{\beta}{2}
                                                      \tilde{{D}}^2 & 0 \\
                           0 & \cos^2 \frac{\beta}{2}{I}_Z + \sin^2 \frac{\beta}{2}
               {D}^2  \end{array} \right)
  \left( \begin{array}{cc} \tilde{{W}} & 0  \\
                              0   &  {V} \end{array} \right)^\dagger
\end{equation}
\end{widetext}
where  $\tilde{{W}}$ of dimension $N\times N$ is a unitary matrix
composed of columns ${W}$:
\begin{equation}
       \tilde{{W}} = \left( {W}, \bar{{W}}
       \right) \quad \mbox{where} \quad
  \tilde{{W}} \tilde{{W}}^\dagger =
  \tilde{{W}}^\dagger  \tilde{{W}} =
  {I}_N,
\end{equation}
and $\bar{{W}}$ is the unitary complement.

The $\tilde{{D}}$ matrix, on the other hand, is the
$Z\times Z$  matrix ${D}$ completed to the dimension
$N\times N$  by zeros:
\begin{equation}\label{matrixD}
     \tilde{{D}}    =
   \left( \begin{array}{cc}   {D} &  0  \\
                              0   &  0          \end{array} \right).
\end{equation}

Since the first and third matrices on the right hand side of Eq.~(\ref{oo-matrix})
are unitary and the second
matrix is diagonal, the inverse of the overlap matrix reads:
\begin{widetext}
\begin{equation}
{\tilde {O}}(\beta)^{-1}
= \left( \begin{array}{cc} \tilde{{W}} & 0  \\
                           0   &  {V} \end{array} \right)
    \left( \begin{array}{cc}  \displaystyle
  \frac{1}{\cos^2 \frac{\beta}{2}{I}_N + \sin^2 \frac{\beta}{2}
                          \tilde{{D}}^2} & 0 \\
  0 &  \displaystyle
    \frac{1}{\cos^2 \frac{\beta}{2}{I}_N + \sin^2 \frac{\beta}{2}
                                     {D}^2  } \end{array} \right)
    \left( \begin{array}{cc} \tilde{{W}} & 0  \\
                              0   &  {V} \end{array} \right)^\dagger
                              {\tilde {O}}(- \beta) .
\end{equation}
\end{widetext}
This expression  shows explicitly
that the inverse of the overlap matrix
${\tilde O}(\beta)^{-1}_{ij}$ can be singular only if  $\beta =\pi$.
Indeed, in the case of
$N$=$Z$, the $i$th denominator becomes zero,
\begin{equation}
\left( \cos^2 \frac{\beta}{2}{I}_N + \sin^2 \frac{\beta}{2}
                                     \tilde{{D}}^2 \right)_i = 0 ,
\end{equation}
only when $\beta = \pi $ and the corresponding
singular value vanishes: $D_i$=0. In the case of $N\ne Z$, the only difference
is that the matrix $\tilde{{D}}$ contains zeros by construction, see
Eq.~(\ref{matrixD}). Fortunately, for the isospin projection, the singularity at
$\beta = \pi$ is compensated by the Jacobian
$\; \sin\beta \;d\beta \;$.

\subsection{Isospin structure of the density matrix}

To calculate the transition density matrix and determine
its isotopic structure
it is convenient to introduce auxiliary ket-states
defined as:
\begin{equation}\label{eqn:nort-st}
\tilde{\phi}_i({\boldsymbol x}, q, \beta) \equiv
\sum_{j=1}^A\tilde{\varphi}_j({\boldsymbol x}, q, \beta)\; {\tilde O}(\beta)^{-1}_{ji}.
\end{equation}
where the space-spin coordinates are abbreviated as
${\boldsymbol x} = ({\boldsymbol r},\sigma )$.
This allows for rewriting the transition
density matrix into a diagonal form:
\begin{equation}\label{eqn:diag-den}
\tilde{\rho}({\boldsymbol x}, q; {\boldsymbol x}', q') \equiv \sum_{i=1}^A
\varphi^*_i({\boldsymbol x}', q')\;
                           \tilde{\phi}_i({\boldsymbol x}, q, \beta ),
\end{equation}
where the summation runs over a set of s.p. states, in full analogy with
the HF particle density matrix. To alleviate the notation, in what follows
we do not indicate the dependence of the transition density
matrices on the isorotation angle $\beta$.

An arbitrary matrix  in the isospace can be decomposed in terms of the isospin Pauli
matrices. In particular, the
transition density matrix $\tilde \rho ({\boldsymbol x}, q; {\boldsymbol x}', q')$ can be expressed as:
\begin{eqnarray}
\tilde \rho ({\boldsymbol x}, q; {\boldsymbol x}', q') & = &
\frac{1}{2} \tilde \rho_{00} ({\boldsymbol x}, {\boldsymbol x}') \delta_{q q'} \nonumber \\
& + & \frac{1}{2} \sum_{\alpha = 0,\pm 1} \tau_{1\alpha} (q, q')
\tilde \rho_{1 \alpha} ({\boldsymbol x}, {\boldsymbol x}'),
\end{eqnarray}
where
\begin{eqnarray}
\tau_{10} (q,q') = \langle q | \hat \tau_{10} | q' \rangle = q \delta_{q,q'} , \\
\tau_{1\pm 1} (q,q') =  \langle q | \hat \tau_{1\pm 1} | q' \rangle  =
\mp \sqrt2 \delta_{q, \pm 1} \delta_{q' , \mp 1}.
\end{eqnarray}
The isoscalar $\tilde \rho_{00}$ and isovector  $\tilde \rho_{1\alpha}$
transition density matrices are calculated by contracting  $\tilde \rho$:
\begin{subequations}
\label{isorho}
\begin{eqnarray}
\tilde \rho_{00}
   & = & \mbox{Tr}\; \tilde \rho ,\\
\tilde \rho_{10}   & = & \mbox{Tr}\; ( \tau_{10} \tilde \rho ) ,\\
\tilde \rho_{1\pm 1} & = & - \mbox{Tr}\; ( \tau_{1\mp 1} \tilde \rho ).
\end{eqnarray}
\end{subequations}

The use of the  two-dimensional spinor notation  is  convenient for
analytical derivations. However, from the numerical-programming perspective it is
better to present expressions in a one-dimensional
form, because it allows us to exploit the separability of proton and neutron
wave functions and densities -- a common feature of essentially all the
existing HF codes including the HFODD solver used in this work.

To write the isoscalar and isovector densities in the one-dimensional form
we introduce
two sets of neutron-like ($i=1,2,\ldots ,N$) wave functions:
\begin{eqnarray}\label{nn-state}
\tilde{\phi}_i^{(n)} ({\boldsymbol x}) & \equiv &
\sum_{j=1}^N \varphi_j^{(n)} ({\boldsymbol x})\; {\tilde O}(\beta)^{-1}_{ji}, \\
\label{np-state} \tilde{\phi}_i^{(p)} ({\boldsymbol x}) & \equiv &
\sum_{j=N+1}^{A} \varphi_j^{(p)} ({\boldsymbol x})\; {\tilde O}(\beta)^{-1}_{ji},
\end{eqnarray}
and two sets of proton-like ($i=N+1,N+2,\ldots ,A$)
 wave functions:
\begin{eqnarray}\label{pnstate}
\tilde{\phi}_i^{(n)} ({\boldsymbol x}) & \equiv &
\sum_{j=1}^N \varphi_j^{(n)} ({\boldsymbol x})\; {\tilde O}(\beta)^{-1}_{ji}, \\
\label{pp-state} \tilde{\phi}_i^{(p)} ({\boldsymbol x}) & \equiv &
\sum_{j=N+1}^{A} \varphi_j^{(p)} ({\boldsymbol x})\; {\tilde O}(\beta)^{-1}_{ji}.
\end{eqnarray}
Again, the dependence of these states on the isorotation angle $\beta$
is not explicitly indicated.
Using these wave functions, we define two neutron-like
densities:
\begin{eqnarray}\label{rho-nn}
           \tilde{\rho}_{n}^{(n)} ({\boldsymbol x},{\boldsymbol x}')& = &
           \sum_{i=1}^N  \varphi_i^{(n) * } ({\boldsymbol x}')\;
            \tilde{\phi}_i^{(n)} ({\boldsymbol x})   \\ \label{rho-np}
           \tilde{\rho}_{n}^{(p)} ({\boldsymbol x},{\boldsymbol x}')& = &
           \sum_{i=1}^N  \varphi_i^{(n) * } ({\boldsymbol x}')\;
            \tilde{\phi}_i^{(p)} ({\boldsymbol x}),
\end{eqnarray}
and two proton-like densities:
\begin{eqnarray} \label{rho-pn}
           \tilde{\rho}_{p}^{(n)}  ({\boldsymbol x},{\boldsymbol x}') & = &
           \sum_{i=N+1}^A  \varphi_i^{(p) * } ({\boldsymbol x}')\;
            \tilde{\phi}_i^{(n)} ({\boldsymbol x})  \\ \label{rho-pp}
           \tilde{\rho}_{p}^{(p)} ({\boldsymbol x},{\boldsymbol x}')& = &
           \sum_{i=N+1}^A  \varphi_i^{(p) * } ({\boldsymbol x}')\;
            \tilde{\phi}_i^{(p)} ({\boldsymbol x}).
\end{eqnarray}
In these definitions, subscripts indicate whether the density is neutron- or
proton-like and  superscripts indicate whether  the
summation indices  in Eqs.~(\ref{nn-state})--(\ref{pp-state}) run over neutron or proton states.
The densities $\tilde{\rho}_{n}^{(p)} ({\boldsymbol x},{\boldsymbol x}')$
and $\tilde{\rho}_{p}^{(n)}  ({\boldsymbol x},{\boldsymbol x}')$ can be associated, respectively,  with the raising  and lowering  components of the
analog spin \cite{[Mac66],[Mek68]}.

The isoscalar and isovector densities (\ref{isorho})
can now be expressed by using
the auxiliary densities (\ref{rho-nn})--(\ref{rho-pp}), and they read
\begin{eqnarray}
     \tilde{\rho}_{00}  ({\boldsymbol x},{\boldsymbol x}')
    & = &\cos\frac{\beta}{2} \left( \tilde{\rho}_{n}^{(n)}  ({\boldsymbol x},{\boldsymbol x}')
                                 + \tilde{\rho}_{p}^{(p)}  ({\boldsymbol x},{\boldsymbol x}') \right)
                                 \nonumber \\
    & - &\sin\frac{\beta}{2} \left( \tilde{\rho}_{n}^{(p)}  ({\boldsymbol x},{\boldsymbol x}')
                                 -\tilde{\rho}_{p}^{(n)}  ({\boldsymbol x},{\boldsymbol x}') \right) , \\
     \tilde{\rho}_{10}  ({\boldsymbol x},{\boldsymbol x}')
    & = &\cos\frac{\beta}{2} \left( \tilde{\rho}_{n}^{(n)}  ({\boldsymbol x},{\boldsymbol x}')
                                 - \tilde{\rho}_{p}^{(p)}  ({\boldsymbol x},{\boldsymbol x}') \right)
                                 \nonumber \\
    & - &\sin\frac{\beta}{2} \left( \tilde{\rho}_{n}^{(p)}  ({\boldsymbol x},{\boldsymbol x}')
                                 +\tilde{\rho}_{p}^{(n)}  ({\boldsymbol x},{\boldsymbol x}') \right) ,
\end{eqnarray}
\begin{eqnarray}
     \tilde{\rho}_{11}  ({\boldsymbol x},{\boldsymbol x}')
     =  - & \sqrt{2} & \cos\frac{\beta}{2}  \tilde{\rho}_{p}^{(n)}  ({\boldsymbol x},{\boldsymbol x}')
     \nonumber \\
        + & \sqrt{2} & \sin\frac{\beta}{2}  \tilde{\rho}_{p}^{(p)} ({\boldsymbol x},{\boldsymbol x}') ,
                          \\
      \tilde{\rho}_{1-1}  ({\boldsymbol x},{\boldsymbol x}')
     =   & \sqrt{2} &
    \cos\frac{\beta}{2}  \tilde{\rho}_{n}^{(p)}  ({\boldsymbol x},{\boldsymbol x}')  \nonumber  \\
  + &\sqrt{2} & \sin\frac{\beta}{2}  \tilde{\rho}_{n}^{(n)}  ({\boldsymbol x},{\boldsymbol x}') .
                                 \end{eqnarray}

\subsection{Hamiltonian kernel}

Conventional Skyrme EDF
can be expressed by bilinear forms of six isoscalar ($t=0$)
and six  isovector ($t=1$) local densities, including the particle $\rho_t$,
kinetic $\tau_\tau$, spin $\vec{s}_t$, spin-kinetic $\vec{T}_t$, current $\vec{j}_t$,
and spin-current $\tensor{J}_t$ densities and
their derivatives. Standard definitions of these densities can be found in numerous
references, see, e.g., Refs.~\cite{[Ben03],[Per04]} and references quoted therein.
It is to be noted that in the standard MF
theory the proton and neutron s.p. wave functions are not mixed, i.e., the proton-neutron symmetry is strictly conserved \cite{[Per04],[Roh09]}.
Therefore,  the MF isoscalar (isovector) densities are
simply sums (differences) of neutron and proton densities since only the third component of the isovector density
is nonzero.

The isospin projection essentially preserves the functional form of the Skyrme EDF derived by
averaging the isospin-invariant Skyrme interaction over the Slater determinant.
All what needs to be done is a replacement of the local density matrix by the
corresponding transition density matrix. Moreover, the
bilinear terms that depend of the isovector densities must be
replaced by the full isoscalar products of the corresponding
isovector transition densities.
Special care should be taken of the density-dependent term of the Skyrme
interaction as  the extension  of this term
to the transition case
is undefined in the process of
averaging the Skyrme interaction over the Slater determinant. Following the argumentation of
Refs.~\cite{[Ang01],[Rob07]}, we replace the isoscalar density  by the
transition isoscalar density matrix, that is, $\rho_{n}+\rho_{p}
\longrightarrow \tilde \rho_{00}$.

As for the Coulomb-interaction kernel, it depends on the isoscalar
and isovector transition densities in the following way:
\begin{eqnarray}
 & \langle\Phi | \hat V^C \hat{R}(\beta)|\Phi\rangle = &\nonumber \\
 & \displaystyle = \frac{e^2}{4} \int d \vec{r}_1 \int d \vec{r}_2
       \frac{1}{r_{12}}\sum_{\lambda = 0}^2 \sum_{\mu = - \lambda}^{\lambda}
        {\cal V}^{C}_{\lambda \mu }( \vec{r}_1, \vec{r}_2 ), &
\end{eqnarray}
where
\begin{widetext}
\begin{eqnarray}\label{VC00c}
       {\cal V}^{C}_{00} & = &
       \frac{1}{2} \left\{ \tilde\rho_{00} (\vec{r}_1 )  \tilde\rho_{00} (\vec{r}_2 )
       + \frac{1}{3} \tilde\rho_{1} (\vec{r}_1 ) \circ \tilde\rho_{1} (\vec{r}_2 ) \right\}  \nonumber \\
           &   - &
         \frac{1}{4} \left\{ \tilde\rho_{00} (\vec{r}_2 , \vec{r}_1 )  \tilde\rho_{00} (\vec{r}_1, \vec{r}_2 )
       + \frac{1}{3} \tilde\rho_{1} (\vec{r}_2, \vec{r}_1 ) \circ \tilde\rho_{1} (\vec{r}_1, \vec{r}_2 ) \right. \nonumber \\
      &&+  \left. \tilde{\vec s}_{00} (\vec{r}_2 , \vec{r}_1 ) \cdot \tilde{\vec s}_{00} (\vec{r}_1, \vec{r}_2 )
       +  \frac{1}{3} \tilde{\vec s}_{1} (\vec{r}_2, \vec{r}_1 ) \circ \cdot \tilde{\vec s}_{1} (\vec{r}_1, \vec{r}_2 )
       \right\} , \\
\label{VC10c}
       {\cal V}^{C}_{10}  &=&  - \tilde\rho_{00} (\vec{r}_1 )  \tilde\rho_{10} (\vec{r}_2 )
     + \frac{1}{2} \left\{ \tilde\rho_{00} (\vec{r}_2, \vec{r}_1 )  \tilde\rho_{10} (\vec{r}_1, \vec{r}_2 )
            + \tilde{\vec s}_{00} (\vec{r}_2, \vec{r}_1 ) \cdot \tilde{\vec s}_{10} (\vec{r}_1, \vec{r}_2 )
     \right\} , \\
\label{VC11c}
       {\cal V}^{C}_{1\pm 1}  &=&   \tilde\rho_{00} (\vec{r}_1 )  \tilde\rho_{1\mp 1} (\vec{r}_2 )
     - \frac{1}{2} \left\{ \tilde\rho_{00} (\vec{r}_2, \vec{r}_1 )  \tilde\rho_{1\mp 1} (\vec{r}_1, \vec{r}_2 )
     +                \tilde{\vec s}_{00} (\vec{r}_2, \vec{r}_1 ) \cdot \tilde{\vec s}_{1\mp 1} (\vec{r}_1, \vec{r}_2 )
     \right\} , \\
\label{VC20c}
       {\cal V}^{C}_{20} & = &
       \frac{1}{3} \left\{  \tilde\rho_{10}  (\vec{r}_1 )  \tilde\rho_{1 0} (\vec{r}_2 )
                              + \tilde\rho_{11} (\vec{r}_1 )  \tilde\rho_{1-1} (\vec{r}_2 ) \right\}
       \nonumber \\
  & - &  \frac{1}{6} \left\{ \tilde\rho_{10} (\vec{r}_2, \vec{r}_1 )  \tilde\rho_{1 0} (\vec{r}_1, \vec{r}_2 )
    +  \tilde\rho_{11} (\vec{r}_2, \vec{r}_1 )  \tilde\rho_{1-1} (\vec{r}_1, \vec{r}_2 )
       \right. \nonumber \\
  &&+  \left. \tilde{\vec s}_{10} (\vec{r}_2, \vec{r}_1 ) \cdot \tilde{\vec s}_{1 0} (\vec{r}_1, \vec{r}_2 )
    +  \tilde{\vec s}_{11} (\vec{r}_2, \vec{r}_1 ) \cdot \tilde{\vec s}_{1-1} (\vec{r}_1, \vec{r}_2 )
     \right\} , \\
\label{VC21c}
       {\cal V}^{C}_{2\pm 1}  &=&  -\frac{1}{\sqrt{3}}
        \tilde\rho_{10} (\vec{r}_1 )  \tilde\rho_{1\mp 1} (\vec{r}_2 )
     + \frac{1}{2\sqrt{3}}
               \left\{ \tilde\rho_{10} (\vec{r}_2, \vec{r}_1 )  \tilde\rho_{1\mp 1} (\vec{r}_1, \vec{r}_2 )
     +        \tilde{\vec s}_{10} (\vec{r}_2, \vec{r}_1 ) \cdot \tilde{\vec s}_{1\mp 1} (\vec{r}_1, \vec{r}_2 )
     \right\} , \\
\label{VC22}
       {\cal V}^{C}_{2\pm 2}  &=&  \frac{1}{\sqrt{6}}
        \tilde\rho_{1\mp 1} (\vec{r}_1 )  \tilde\rho_{1\mp 1} (\vec{r}_2 )
     - \frac{1}{2\sqrt{6}} \left\{ \tilde\rho_{1\mp 1} (\vec{r}_2, \vec{r}_1 )  \tilde\rho_{1\mp 1} (\vec{r}_1, \vec{r}_2 )
     +   \tilde{\vec s}_{1\mp 1} (\vec{r}_2, \vec{r}_1 ) \cdot \tilde{\vec s}_{1\mp 1} (\vec{r}_1, \vec{r}_2 )
     \right\} .
\end{eqnarray}
\end{widetext}
Here, the symbols $\cdot$ and $\circ$ stand for the scalar products of
vectors and isovectors, respectively.

\subsection{Coulomb rediagonalization and the isospin mixing}

To properly account for  the isospin mixing effects, following
Refs.~\cite{[Raf09d],[Sat09a]},
the total Hamiltonian
${\hat H}_{NN}$  (\ref{ham}) (strong interaction plus the Coulomb
interaction) is rediagonalized  in the space spanned by the
good-isospin wave functions (\ref{eqn:imk}), and the resulting
eigenstates are denoted by
\begin{equation}\label{mix2}
|n,T_z\rangle
= \sum_{T\geq |T_z|}a^n_{T,T_z}|T,T_z\rangle
\end{equation}
and
numbered by index $n$.
The amplitudes  $a^n_{T,T_z}$ define the degree of
isospin mixing. In particular, the isospin-mixing
parameter for the lowest energy solution $E_{n=1,T_z}$ is defined as
$\alpha_C = 1- |a^{n=1}_{|T_z|,T_z}|^2$.

\begin{figure}[tbp!]
\includegraphics[width=0.9\columnwidth,clip]{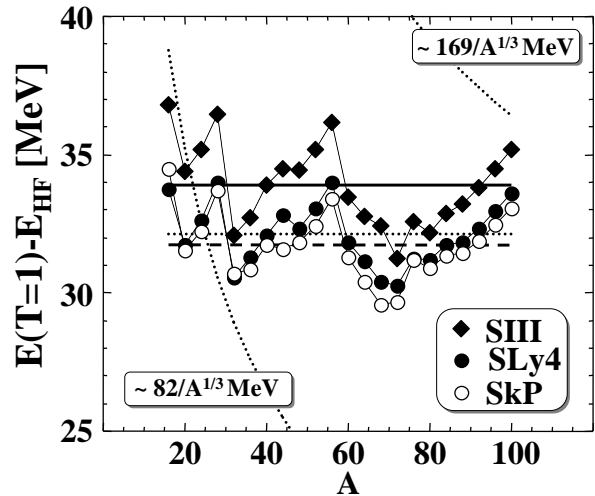}
\caption{Excitation energies of the doorway $T$=1 states in the e-e $N$=$Z$
nuclei relative to the SHF g.s.\ energies $E_{HF}$.
Diamonds, dots, and circles show the self-consistent results
obtained in the isospin-projected DFT  with  SIII~\cite{[Bei75]},
SLy4~\cite{[Cha97]}, and SkP~\cite{[Dob84]} Skyrme functionals,
respectively. Horizontal lines mark the mean DFT values.
Thick dotted lines show estimates based on the
perturbation theory, $E_{T=1}\approx
82/A^{1/3}$\,MeV~\cite{[Sli65],[Boh67]}, and on the hydrodynamical model,
$E_{T=1}\approx 169/A^{1/3}$\,MeV \cite{[Boh67]}.
}\label{doorway}
\end{figure}

Precise determination of the Coulomb mixing constitutes a notoriously
difficult problem, but is strongly motivated by its impact on
fundamental physics tested through the super-allowed $\beta$
decays~\cite{[Har05b],[Tow08],[Mil08]}. Of importance here is to capture the
proper balance between the short-range strong interaction and the
long-range Coulomb polarization. This balance is naturally taken into
account by the DFT approach, but is
not accessible within a perturbative-analysis theory~\cite{[Sli65]}
or hydrodynamical model~\cite{[Boh67]}.
This is illustrated in Fig.~\ref{doorway} that shows   the
excitation energies of the doorway $T$=1 states in e-e $N$=$Z$ nuclei,
i.e., the energies of the $n=2$ states in Eq.~(\ref{mix2})
relative to the HF g.s.\ energies $E_{HF}$.
The estimates given by the perturbative/hydrodynamical approaches
are clearly   at variance with the self-consistent
results.

Figure~\ref{doorway}  shows that not only the values but also the $A$-dependence
of the doorway excitation energies differ substantially
from the self-consistent results, pointing very clearly to a
non-perturbative origin of the Coulomb mixing.
One can also notice that the mean excitation
energies of the doorway states
(directly impacting  the Coulomb mixing)  strongly depend on the EDF
parameterization. The currently used  Skyrme functionals
are not sufficiently  constrained in the isospin sector
to provide reliable estimates of $\alpha_C$. At this point
is also not at all obvious what parts of EDF have to be
refined in order to improve on this situation, see Ref.~\cite{[Sat09a]}
for further details.

\section{Illustrative examples}\label{applications}

All calculations presented in this work were obtained by using
isospin projection implemented in the DFT solver
HFODD~\cite{[Dob09e],[Dob09f]}. We used the harmonic-oscillator basis of
$N_0$=12 shells. We considered
SIII~\cite{[Bei75]}, SLy4~\cite{[Cha97]}, and SkP~\cite{[Dob84]}
Skyrme functionals, as well as for their Landau and tensor versions
denoted by subscripts $L$ and $T$, respectively, see
Ref.~\cite{[Zal08]} for details. Pairing was neglected.

\subsection{Superdeformed bands in $^{56}$Ni}\label{ni56}

\begin{figure}[tbp!]
\includegraphics[width=\columnwidth,clip]{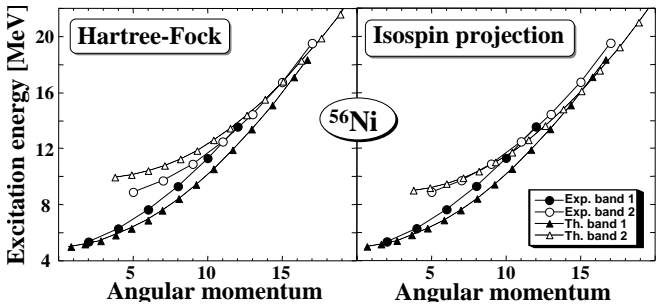}
\caption{Superdeformed bands in $^{56}$Ni. Experimental data
for Band 1 (dots) and Band 2 (circles) of Ref.~\cite{[Rud99fw]} are
compared with HF+SLy4$_L$ predictions  marked by full and open
triangles, respectively. The left   panel shows the standard HF results
while the predicted $T$=0 SD bands obtained in the isospin-projected
approach  are displayed in the right panel.
}\label{fig3}
\end{figure}

We begin discussion from the case of SD
bands  in
$^{56}$Ni. As discussed in Sec.~\ref{oddodd}, the standard cranked HF theory gives a reasonable reproduction of Band 1 while there are problems with theoretical interpretation of Band 2 built on p-h excitations to the  [440]1/2 proton and neutron levels.
The results are summarized in
Fig.~\ref{fig3}.
The HF (left panel) and isospin-projected (right panel)
calculations were performed with SLy4$_L$  functional. Since the HF configuration
corresponding to Band 1 is predominantly isospin symmetric, it is
only weakly affected by the isospin projection. The calculated
isospin impurity within this band is small ($\sim 2$\%). This value is similar to that obtained for the
spherical g.s.\ configuration of $^{56}$Ni. On the other hand, the
Band 2 based on the $\pi$[440]1/2 orbital represents, before
rediagonalization, almost  equal mixture of the $T$=0
and $T$=1 components. Therefore, its isospin impurity assumes an
unrealistic value of about 50\%.

Following the  isospin projection, the low-spin part of the $T$=0
SD band originating from Band 2  is shifted down by
about 0.95\,MeV with respect to the unprojected HF result.
The isospin impurity in this band is in the range of 6\%
to 8\%, and it slowly increases as a function of the angular momentum. In
fact, these values may indicate a presence of an uncontrolled isospin
mixing related, most likely, to the angular-momentum
non-conservation.
As shown schematically in Fig.~\ref{fig2}, a  $T$=1 SD partner structure is expected to lie higher in energy. We indeed
calculate the $T$=1 band (not shown in Fig.~\ref{fig3})  to lie about 2.5\,MeV above the $T$=0 band. As expected, the $T$=0 and $T$=1
bands projected from the $\nu$[440]1/2 MF configuration are almost
identical to those projected from the $\pi$[440]1/2 MF configuration.

By comparing the results of isospin-projected calculations with experiment, we see a significant improvement as compared to standard HF.
The $T$=0 Band 2 agrees  well with experiment, both in terms of
excitation energy and  MoI. The predicted crossing between $T$=0
Bands 1 and 2 occurs
around spin
14\,$\hbar$, i.e., $\sim 2\,\hbar$ too high as compared to the data.
This discrepancy, however, rather reflects a deficiency of our model
in describing the MoI of Band 1, which is slightly overestimated as our calculations neglect pairing correlations that are expected to be important in this configuration \cite{[Rud99fw]}. The large energy splitting between
the $T$=0  and  $T$=1 Band 2 doublet
is consistent with observation of only one SD side band.
In short, the isospin-projected MF
theory provides a quantitative explanation of experimental data
on collective band structures in $^{56}$Ni.

\subsection{Isospin symmetry breaking at band termination}\label{term}

Terminating states or seniority isomers are fully-stretched p-h configurations
with a  maximum-spin  $I_{max}$ that
can be built within a given SM space of valence particles. Because of their
simple SM  character, terminating states
provide a  robust probe of  SM and MF theories and corresponding  effective
interactions.
In this context, of particular interest are the terminating states associated with
the $[f_{7/2}^n]_{I_{max}}$ and  $[d_{3/2}^{-1} f_{7/2}^{n+1}]_{I_{max}}$ configurations
(in the following, $n$ denotes  a number of valence particles outside the $^{40}$Ca core)
in $20\leq Z\leq N \leq 24$ nuclei from the lower$-fp$ shell
($A\sim 44$), which were systematically measured during the last decade
\cite{[Len99w],[Bed01w],[Bra01w],[Lac03w],[Lac05w],[Chi07w]}.

According to MF calculations, these specific states appear to have
almost spherical shapes; hence, the
correlations resulting from the angular-momentum
restoration are practically negligible there~\cite{[Zdu08]}. Hence, they can be regarded as extreme
cases of an almost undisturbed s.p.\ motion, thus offering an  excellent
playground to study, among others,  time-odd densities and fields,
spin-orbit force~\cite{[Zdu05yw],[Sto06b],[Zal07aw]}, tensor
interactions~\cite{[Sat08]}, and the isospin dependence
of cross-shell ($sd-fp$) p-h matrix elements~\cite{[Sto06b],[Sat07w]}.

While most of the terminating states are uniquely defined,  the $[d_{3/2}^{-1} f_{7/2}^{n+1}]_{I_{max}}$ states in
the $N$=$Z$ nuclei provide a notable exception. Indeed, within the MF approximation these particular
states can be created by promoting either one proton, $[d_{3/2}^{-1}
f_{7/2}^{n+1}]_{I_{max}}^\pi$, or one neutron, $[d_{3/2}^{-1}
f_{7/2}^{n+1}]_{I_{max}}^\nu$, across the magic gap 20, Because the Coulomb energy difference between these configurations is small, the resulting energy
levels are almost degenerate:
$E([d_{3/2}^{-1} f_{7/2}^{n+1}]_{I_{max}}^\pi)
\approx E([d_{3/2}^{-1} f_{7/2}^{n+1}]_{I_{max}}^\nu)$.
Therefore,  we encounter  problems related to isospin symmetry
similar to those
discussed earlier in Sec.~\ref{oddodd}.

As in the case of $^{56}$Ni, we encounter the situation where
the isospin symmetry is manifestly broken by the MF approximation
and the  predictions are at variance with empirical data.
The difficulty with the isospin content of terminating states was  recognized in Ref.~\cite{[Sto06b]}, where a
purely phenomenological method of isospin restoration was
 proposed. It has resulted in a good MF description of
experimental data, at the level  of the state-of-the-art SM calculations.
Quantitatively, however, the estimated energy
correction due to the isospin projection appeared to
have surprisingly strong $A$-dependence, changing quite rapidly from
$\delta E_T$$\approx$2\,MeV in $^{40}$Ca  down to
$\delta E_T$$\approx$1\,MeV in $^{46}$V, see Fig.~4 of Ref.~\cite{[Sto06b]}. This trend has been
found to depend  weakly  on the EDF parameterization.

\begin{figure}[tbp!]
\includegraphics[width=0.9\columnwidth]{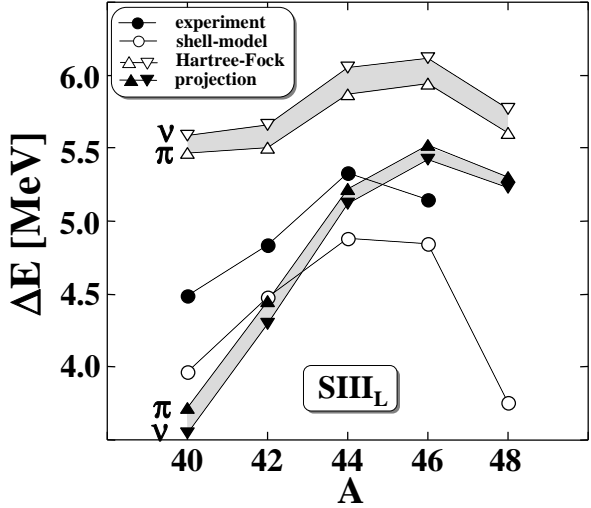}
\caption{Energy differences $\Delta E$ (\protect\ref{delE})
between the terminating states in Ca,Sc,Ti,and Cr $N$=$Z$ nuclei.
Open and full triangles mark, respectively,
standard and isospin-projected HF results
for the
$[d_{3/2}^{-1} f_{7/2}^{n+1}]_{I_{max}}^\pi$ (up triangles) and
$[d_{3/2}^{-1} f_{7/2}^{n+1}]_{I_{max}}^\nu$ (down triangles)
configurations.
Calculations were performed by using the SIII$_L$ functional~\cite{[Zal08]}.
Experimental data (dots) and
SM results of Ref.~\cite{[Sto06b]} (circles) are shown for comparison.
Note the limited energy scale.
}\label{termin}
\end{figure}

In the present work, we repeat calculations of Ref.~\cite{[Sto06b]},
however, by using the
mathematically rigorous isospin projection of
Sec.~\ref{theory}. The energies of the isospin-projected
states $[d_{3/2}^{-1} f_{7/2}^{n+1}]_{I_{max}}^{T=0}$
are shown in Fig.~\ref{termin} relative to those of the $[f_{7/2}^n]_{I_{max}}$
configurations:
\begin{equation}
\label{delE}
\Delta E =
E([d_{3/2}^{-1} f_{7/2}^{n+1}]_{I_{max}})- E([f_{7/2}^n]_{I_{max}}) .
\end{equation}
Full up and down triangles show results obtained by projecting the
$T$=0 component of the Slater determinant corresponding to either one
proton, $|[d_{3/2}^{-1} f_{7/2}^{n+1}]_{I_{max}}^\pi \rangle$, or one
neutron, $|[d_{3/2}^{-1} f_{7/2}^{n+1}]_{I_{max}}^\nu \rangle$, p-h
excitation through the magic gap 20, respectively. For comparison,
the unprojected HF energies for the proton and neutron configurations
are shown with the open up and down triangles, respectively. Here,
all calculations were performed by using the SIII~\cite{[Bei75]} Skyrme
functional.

The isospin-projected results obtained from the $|[d_{3/2}^{-1}
f_{7/2}^{n+1}]_{I_{max}}^\pi \rangle$ and $|[d_{3/2}^{-1}
f_{7/2}^{n+1}]_{I_{max}}^\nu \rangle$ HF configurations are similar but
not identical, reflecting small polarization differences   due to
proton and neutron p-h excitations. Note, that
slightly better results are obtained by projecting from the proton
configurations as the resulting levels appear slightly lower in energy.
This is probably not surprising, because these states
include directly the polarization of the Coulomb field by the proton
p-h excitation and thus they have slightly richer isospin structure
than their neutron counterparts.

\begin{figure}[tbp!]
\includegraphics[width=0.9\columnwidth,clip]{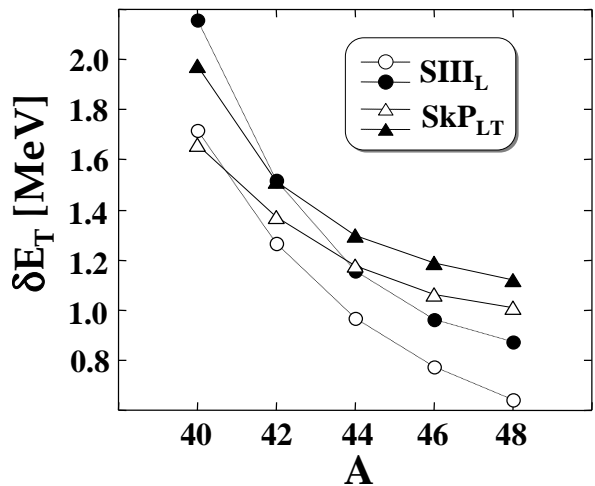}
\caption{Full symbols label energy correction $\delta E_T$
(\protect\ref{isocor}) calculated by projecting good isospin from
the HF states corresponding to the
$[d_{3/2}^{-1} f_{7/2}^{n+1}]_{I_{max}}^\pi$ configurations in $N$=$Z$ nuclei
and subsequent Coulomb rediagonalization. Open symbols represent
results of a phenomenological method proposed in
Ref.~\protect\cite{[Sto06b]}.
Calculations were carried out by using the SIII$_L$ (dots) and SkP$_{LT}$
(triangles) Skyrme functionals~\cite{[Zal08]}.
See text for further details.
}\label{DET}
\end{figure}

The results of rigorous isospin projection closely follow those obtained
in the
phenomenological approach of Ref.~\cite{[Sto06b]}. As shown in
Fig.~\ref{DET}, the energy corrections
calculated in the isospin-projected HF,
\begin{equation}
\label{isocor}
\delta E_T = \tfrac{1}{2}
 \left[E([d_{3/2}^{-1} f_{7/2}^{n+1}]_{I_{max}}^{T=1})
      -E([d_{3/2}^{-1} f_{7/2}^{n+1}]_{I_{max}}^{T=0}) \right],
\end{equation}
also exhibit an appreciable $A$-dependence, in a close analogy to the previously obtained
phenomenological trend. Consequently, the
$A$-dependence of  $\Delta E$ (\ref{delE}),
calculated using the isospin projected EDF approach also shows a different
pattern than experiment and SM. This result
is  weakly dependent on EDF parameterization, suggesting that there
is some generic problem pertaining to the standard form of the Skyrme EDF.

Recently, we have developed a new class of the Skyrme functionals
with spin-orbit and tensor terms locally refitted, to reproduce the
$f_{7/2}-f_{5/2}$ spin-orbit splitting in $^{40}$Ca, $^{48}$Ca, and
$^{56}$Ni~\cite{[Zal08]}.  The spin-orbit and
the tensor strengths  obtained in this way turned out to be
fairly independent of other coupling constants
of the Skyrme functional. This result
indicates that the strong dependence of the spin-orbit
strength on EDF parameterization, in particular on the isoscalar
effective mass \cite{[Les07]}, is likely to be
an artifact of fitting protocols based
predominantly on data pertaining to bulk nuclear properties.
Indeed, as discussed in Ref.~\cite{[Toi08]}, the
use of inaccurate models in the fitting procedure can lead to
results that strongly depend on the fitting protocol itself; hence,
can result in contradictory information on the key model parameters
(compare, e.g., results of Refs.~\cite{[Les07]} and \cite{[Klu09]}). It
seems that this is exactly the case for the current parameterizations
of the Skyrme EDF. As shown recently in Ref.~\cite{[Ben09aa]},
parameterizations that correctly describe the spin-orbit properties
in light nuclei do not fare well in heavier systems. This again points
to limitations of the second-order Skyrme EDF \cite{[Kor08]}
and to a danger of drawing conclusions on  tensor interactions
from global fits ~\cite{[Les07]}.

Applications of new functionals  to the terminating states in $N\ne Z$,
$A\sim 44$ nuclei~\cite{[Sat08a]} have revealed that removal of the
artificial isoscalar effective mass scaling from the spin-orbit
restores the effective mass scaling in the s.p.\ level
density. As a consequence, only the forces having large isoscalar
effective masses ($\frac{M^*}{M}\geq 0.9$) such as  SkP$_T$ and
SkO$_T$ \cite{[Sat08]}, are able to
reproduce empirical data involving  s.p.\ levels in light nuclei.
\begin{figure}
\includegraphics[width=0.9\columnwidth]{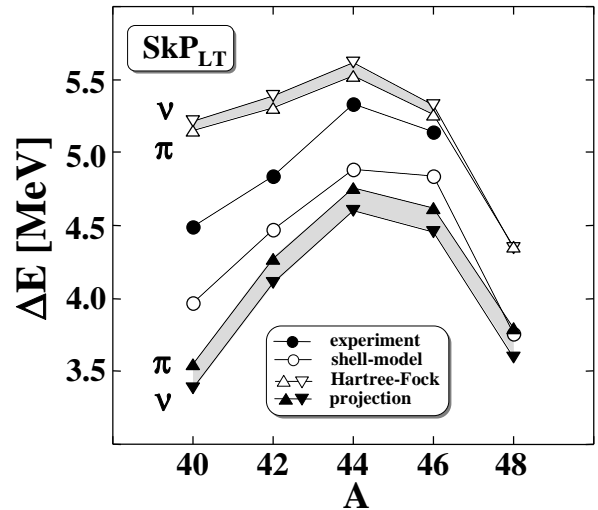}
\caption{Similar to Fig.~\protect\ref{termin} except for the modified
Skyrme functional SkP$_{LT}$ \cite{[Zal08]}.
}\label{fig7}
\end{figure}

This conclusion is nicely corroborated by
results presented in Fig.~\ref{fig7}, which shows the predictions for terminating states using  the new Skyrme parameterization SkP$_{LT}$.
It is rewarding to see that the modified functional yields
results that are very consistent with SM. This is
particularly true for the isospin projection from the $|[d_{3/2}^{-1}
f_{7/2}^{n+1}]_{I_{max}}^\pi\rangle$ configurations. The new values of
 $\Delta E$
reasonably agree with   experiment, considering the energy range of the plot, While the  detailed $A$-dependence is reproduced very well, theoretical  $\Delta E$ curve slightly underestimates experiment. Identification of a
specific source of this remaining discrepancy requires further studies.

\section{Summary and conclusions}
\label{summ}

In this work, we carried out the  theoretical analysis of
isospin breaking in nuclei around $N$=$Z$ based on the density functional theory. We show that the spontaneous breaking of isospin symmetry inherent to  MF  limits the applicability of self-consistent theories, such as HF and DFT, to nuclear states with $T$=0. To remedy this problem, we propose
a new isospin-symmetry restoration scheme based on the rediagonalization
technique in good-isospin basis.

The isospin projection algorithm is described in detail, including the derivation of essential Hamiltonian kernels. We demonstrate that isospin projection is  free from problems plaguing
other symmetry restoration schemes  (particle number, angular momentum)  related to kernel singularities.

Applications of the isospin-projected DFT approach include  the
SD bands in $^{56}$Ni and terminating states in $N$=$Z$ medium-mass
nuclei. In both cases, the symmetry restoration remedies previously noted deficiencies of the HF method and  significantly improves agreement with experiment. The results for terminating states corroborate previous suggestions that large isoscalar effective masses are needed to
reproduce experimental spectroscopic  data involving  s.p. levels.

The examples presented in  this work primarily concern high-spin configurations in medium mass nuclei in which  pairing correlations are expected to be weak.
To be able to address theoretically a variety of low-spin phenomena and observables  around the $N$=$Z$ line, such as ground states of odd-odd nuclei, binding energy staggering, superallowed beta decays, and charge-exchange reactions, isovector and isoscalar pairing correlations must be incorporated and the proton-neutron symmetry of HF must be broken before isospin projection \cite{[Per04],[Roh09]}. Those will be subjects of our forthcoming studies.

This work was supported in part by the Polish Ministry of Science
under Contracts No.~N~N202~328234 and N~N202~239037, Academy of Finland and
University of Jyv\"askyl\"a within the FIDIPRO programme, and by the Office of
Nuclear Physics,  U.S. Department of Energy under Contract Nos.
DE-FG02-96ER40963 (University of Tennessee) and and DE-FC02-07ER41457 (UNEDF
SciDAC Collaboration).


\begin{thebibliography}{10}

\bibitem{[Hoh64w]}
{P. Hohenberg and W. Kohn, Phys. Rev. {\bf 136}, B864 (1964).}

\bibitem{[Koh65w]}
{W. Kohn and L.J. Sham, Phys. Rev. {\bf 140}, A1133 (1965).}

\bibitem{[Lev79]}
{M. Levy. Proc. Nat. Acad. Sci. {\bf 76}, 6062 (1979)}.

\bibitem{[Lev82]}
{M. Levy. Phys. Rev. A {\bf 26}, 1200 (1982)}.

\bibitem{[Lie83]}
{E. Lieb, Int. J. Quant. Chem. {\bf 24}, 243 (1983)}.

\bibitem{[Lie85]}
{M. Levy and J. P. Perdew, in {\it Density Functional Methods in Physics},
  edited by R.M. Dreizler and J. da Providencia (Plenum, New York, 1985)}.

\bibitem{[Eng88a]}
{H. Englisch, H. Fieseler, and A. Haufe, Phys. Rev. A {\bf 37}, 4570 (1988)}.

\bibitem{[Gor96a]}
{A. G\"orling, Phys. Rev. A {\bf 54}, 3912 (1996)}.

\bibitem{[Ben03]}
{M. Bender, P.-H. Heenen, and P.-G. Reinhard, Rev. Mod. Phys. {\bf 75}, 121
  (2003)}.

\bibitem{[Lal04]}
{{\it Extended Density Functionals in Nuclear Structure Physics}, ed. by G.A.
  Lalazissis, P. Ring, and D. Vretenar (Springer Verlag, 2004)}.

\bibitem{[Pro05]}
{E. Prodan, J. Phys. A {\bf 38}, 5647 (2005)}.

\bibitem{[Eng07]}
{J. Engel, Phys. Rev. C {\bf 75}, 014306 (2007)}.

\bibitem{[Gir08]}
{B.G. Giraud, Phys. Rev. {\bf C77}, 014311 (2008)}.

\bibitem{[Gir08a]}
{B.G. Giraud, B.K. Jennings, and B.R. Barrett, Phys. Rev. {\bf A78}, 032507
  (2008)}.

\bibitem{[Mes09a]}
{J. Messud, M. Bender, and E. Suraud, Phys. Rev. C, in press; arXiv:0904.0162}.

\bibitem{[Sky56xw]}
{T.H.R. Skyrme, Phil. Mag. {\bf 1} (1956) 1043; Nucl. Phys. {\bf 9} (1959)
  615.}

\bibitem{[Car08]}
{B.G. Carlsson, J. Dobaczewski, and M. Kortelainen, Phys. Rev. C {\bf 78},
  044326 (2008)}.

\bibitem{[Wil69]}
{{\it Isospin in nuclear physics}, ed. D.H. Wilkinson (North Holland,
  Amsterdam, 1969)}.

\bibitem{[War06a]}
{D.D. Warner, M.A. Bentley, and P. Van Isacker, Nature Physics {\bf 2}, 311
  (2006)}.

\bibitem{[Rud99fw]}
{ D.~Rudolph, C.~Baktash, M.J.~Brinkman, E. Caurier, D.J.~Dean, M.~Devlin,
  J.~Dobaczewski, P.-H. Heenen, H.-Q.~Jin, D.R.~LaFosse, W. Nazarewicz, F.
  Nowacki, A. Poves, L.L. Riedinger, D.G.~Sarantites, W.~Satu{\l}a, and
  C.H.~Yu, Phys. Rev. Lett. {\bf 82} (1999) 3763.}

\bibitem{[Zal07]}
{M. Zalewski, W. Satu{\l}a, W. Nazarewicz, G. Stoitcheva, and H. Zdu\'nczuk,
  Phys. Rev. C {\bf 75}, 054306 (2007)}.

\bibitem{[Dob07d]}
{J. Dobaczewski, M.V. Stoitsov, W. Nazarewicz, and P.-G. Reinhard, C {\bf 76},
  054315 (2007)}.

\bibitem{[Lac09]}
{D. Lacroix, T. Duguet, and M. Bender, Phys. Rev. C {\bf 79}, 044318 (2009)}.

\bibitem{[Ots98w]}
{T. Otsuka, M. Honma, and T. Mizusaki, Phys. Rev. Lett. {\bf 81} (1998) 1588.}

\bibitem{[Miz00]}
{T. Mizusaki, T. Otsuka, M. Honma, and B.A. Brown, Phys. Scr. {\bf T88}, 107
  (2000)}.

\bibitem{[Cau80]}
{E. Caurier, A. Poves, and A. Zucker, Phys. Lett. {\bf 96B}, 11 (1980); Phys.
  Lett. {\bf 96B}, 15 (1980)}.

\bibitem{[Cau82]}
{E. Caurier and A. Poves, Nucl. Phys. {\bf A385}, 407 (1982)}.

\bibitem{[Raf09d]}
{M. Rafalski, W. Satu{\l}a, and J. Dobaczewski, Int. J. Mod. Phys. {\bf E18},
  958 (2009); arXiv:0811.1135}.

\bibitem{[Sat09a]}
{W. Satu{\l}a, J. Dobaczewski, W. Nazarewicz, and M. Rafalski, Phys. Rev. Lett.
  {\bf 103}, 012502 (2009)}.

\bibitem{[She00a]}
{J.A. Sheikh and P. Ring, Nucl. Phys. A {\bf 665}, 71 (2000)}.

\bibitem{[She02]}
{J.A. Sheikh, P. Ring, E. Lopes, and R. Rossignoli, Phys. Rev. C {\bf 66},
  044318 (2002)}.

\bibitem{[Sto07]}
{M.V. Stoitsov, J. Dobaczewski, R. Kirchner, W. Nazarewicz, and J. Terasaki,
  Phys. Rev. C {\bf 76}, 014308 (2007)}.

\bibitem{[Mar99a]}
{G. Martinez-Pinedo, K. Langanke, and P. Vogel, Nucl. Phys. A {\bf 651}, 379
  (1999)}.

\bibitem{[Vog00]}
{P. Vogel, Nucl. Phys. {\bf A662}, 148 (2000)}.

\bibitem{[Jan05a]}
{J. J\"anecke and T.W. O'Donnell, Phys. Lett. B {\bf 605}, 87 (2005)}.

\bibitem{[Eng70]}
{C.A. Engelbrecht and R.H. Lemmer, Phys. Rev. Lett. {\bf 24}, 607 (1970)}.

\bibitem{[Var88]}
{D.A. Varshalovich, A.N. Moskalev, and V.K. Khersonskii, {\sl Quantum Theory of
  Angular Momentum} (World Scientific, Singapore 1988)}.

\bibitem{[Rod02bw]}
{R. Rodriguez-Guzm\'an, J.L. Egido, and L.M. Robledo, Nucl. Phys. {\bf A709},
  201 (2002).}

\bibitem{[Bla86]}
{J.P.~Blaizot and G.~Ripka, {\it Quantum theory of finite systems\/}, MIT
  Press, Cambridge Mass., 1986}.

\bibitem{[Zdu07]}
{H. Zdu{\'n}czuk, J. Dobaczewski, and W. Satu{\l}a, Int. J. Mod. Phys. E {\bf
  16}, 377 (2007)}.

\bibitem{[Bei75]}
{M. Beiner, H. Flocard, N. Van Giai, and P. Quentin, Nucl. Phys. A {\bf 238},
  29 (1975)}.

\bibitem{[Dob96]}
{J. Dobaczewski, W. Nazarewicz, T.R. Werner, J.-F. Berger, C.R. Chinn, and J.
  Decharg\'e, Phys. Rev. {\bf C53}, 2809 (1996)}.

\bibitem{[Dob09d]}
{J. Dobaczewski, W. Satu{\l}a, B.G. Carlsson, J. Engel, P. Olbratowski, P.
  Powa{\l}owski, M. Sadziak, J. Sarich, N. Schunck, A. Staszczak, M.V.
  Stoitsov, M. Zalewski, and H. Zdu\'nczuk, Comput. Phys. Commun. {\bf 180},
  2361 (2009)}.

\bibitem{[Mac66]}
{W.M. MacDonald, in {\it Isobaric Spin in Nuclear Physics}, Ed. by J.D. Fox and
  D. Robson, Academic Press, 1966}.

\bibitem{[Mek68]}
{A.Z. Mekjian and W.M. MacDonald, Nucl. Phys. A {\bf 121}, 385 (1968)}.

\bibitem{[Per04]}
{E. Perli\'nska, S.G. Rohozi\'nski, J. Dobaczewski, and W. Nazarewicz, Phys.
  Rev. C {\bf 69}, 014316 (2004)}.

\bibitem{[Roh09]}
{S.G. Rohozi\'nski, J. Dobaczewski, and W. Nazarewicz, arXiv:0910.5270}.

\bibitem{[Ang01]}
{M. Anguiano, J.L. Egido, and L.M. Robledo, Nucl. Phys. A {\bf 696}, 467
  (2001)}.

\bibitem{[Rob07]}
{L.M. Robledo, Int. J. Mod. Phys. E {\bf 16}, 337 (2007)}.

\bibitem{[Har05b]}
{J.C. Hardy and I.S. Towner, Phys. Rev. {\bf C71}, 055501 (2005); Phys. Rev.
  Lett. {\bf 94}, 092502 (2005); Phys. Rev. {\bf C79}, 055502 (2009)}.

\bibitem{[Tow08]}
{I.S. Towner and J.C. Hardy, Phys. Rev. C {\bf 77}, 025501 (2008)}.

\bibitem{[Mil08]}
{G.A. Miller and A. Schwenk, Phys. Rev. C {\bf 78}, 035501 (2008)}.

\bibitem{[Sli65]}
{L.A. Sliv and Yu.I. Kharitonov, Phys. Lett. {\bf 16}, 176 (1965)}.

\bibitem{[Boh67]}
{A. Bohr, J. Damg{\aa}rd, and B. Mottelson, in {\it Nuclear Structure}, ed. by
  A. Hossian {\it et al.} (North-Holland Publ. Co., Amsterdam, 1967).}

\bibitem{[Cha97]}
{E. Chabanat, P. Bonche, P. Haensel, J. Meyer, and R. Schaeffer, Nucl. Phys.
  {\bf A627} (1997) 710}.

\bibitem{[Dob84]}
{J. Dobaczewski, H. Flocard and J. Treiner, Nucl. Phys. {\bf A422}, 103
  (1984)}.

\bibitem{[Dob09e]}
{J. Dobaczewski {\it et al.}, Comput. Phys. Commun. {\bf 102}, 166 (1997); {\bf
  102}, 183 (1997); {\bf 131}, 164 (2000); {\bf 158}, 158 (2004); {\bf 167},
  214 (2005); {\bf 180}, 2361 (2009)}.

\bibitem{[Dob09f]}
{J. Dobaczewski, B.G. Carlsson, J. Dudek, J. Engel, P. Olbratowski, P.
  Powa{\l}owski, M. Sadziak, J. Sarich, W. Satu{\l}a, N. Schunck, A. Staszczak,
  M.V. Stoitsov, M. Zalewski, and H. Zdu\'nczuk, HFODD (v2.40h): User's Guide,
  arXiv:0909.3626}.

\bibitem{[Zal08]}
{M. Zalewski, J. Dobaczewski, W. Satu{\l}a, and T.R. Werner, Phys. Rev. C {\bf
  77}, 024316 (2008)}.

\bibitem{[Len99w]}
{S.M. Lenzi, D.R. Napoli, C.A. Ur, D. Bazzacco, F. Brandolini, J.A. Cameron, E.
  Caurier, G. de Angelis, M. De Poli, E. Farnea, A. Gadea, S. Hankonen, S.
  Lunardi, G. Mart\'{\i}nez-Pinedo, Zs. Podolyak, A. Poves, C. Rossi Alvarez,
  J. S\'anchez-Solano, and H. Somacal, Phys Rev. {\bf C60}, 021303(R) (1999)}.

\bibitem{[Bed01w]}
{P. Bednarczyk, W. M{\c e}czy\'nski, J. Stycze\'n, J. Gr{\c e}bosz, M. Lach, A.
  Maj, M. Zi{\c e}bli\'nski, N. Kintz, J.C. Merdinger, N. Schulz, J.P. Vivien,
  A. Bracco, J.L. Pedroza, M.B. Smith, and K.M. Spohr, Acta Phys. Pol. {\bf
  B32}, 747 (2001)}.

\bibitem{[Bra01w]}
{F. Brandolini, N.H. Medina, D. Bazzacco, J.A. Cameron, G. de Angelis, A.
  Gadea, N. Menegazzo, and C. Rossi-Alvarez, Nucl. Phys. {\bf A693}, 517
  (2001)}.

\bibitem{[Lac03w]}
{M. Lach, J. Stycze\'n, W. M\c{e}czy\'nski, P. Bednarczyk, A. Bracco, J.
  Gr\c{e}bosz, A. Maj, J.C. Merdinger, N. Schultz, M.B. Smith, K.M. Spohr, J.P.
  Vivien, and M. Zi\c{e}bi\'nski, Eur. Phys. J. {\bf A16}, 309 (2003)}.

\bibitem{[Lac05w]}
{M. Lach, J. Stycze\'n, W. M\c{e}czy\'nski, P. Bednarczyk, A. Bracco, J.
  Gr\c{e}bosz, A. Maj, J.C. Merdinger, N. Schultz, M.B. Smith, K.M. Spohr, and
  M. Zi\c{e}bi\'nski, Eur. Phys. J. {\bf A25}, 1 (2005)}.

\bibitem{[Chi07w]}
{C.J. Chiara, M. Devlin, E. Ideguchi, D.R. LaFosse, F. Lerma, W. Reviol, S.K.
  Ryu, D.G. Sarantites, O.L. Pechenaya, C. Baktash, A. Galindo-Uribarri, M.P.
  Carpenter, R.V.F. Janssens, T. Lauritsen, C.J. Lister, P. Reiter, D.
  Seweryniak, P. Fallon, A. Gorgen, A.O. Macchiavelli, D. Rudolph, G.
  Stoitcheva, and W.E. Ormand, Phys. Rev. {\bf C75}, 054305 (2007); {\bf C75},
  059904(E) (2007)}.

\bibitem{[Zdu08]}
{H. Zdu{\'n}czuk, W. Satu{\l}a, J. Dobaczewski, and M. Kosmulski, Phys. Rev.
  C{\bf 76}, 044304 (2008).}

\bibitem{[Zdu05yw]}
{H. Zdu{\'n}czuk, W. Satu{\l}a, and R. Wyss, Phys. Rev. {\bf C71} (2005)
  024305; Int. J. Mod. Phys. {\bf E14} (2005) 451; W. Satu{\l}a, R. Wyss, and
  H. Zdu\'nczuk, Eur. Phys. J. A {\bf 25}, s01, (2005) 551. }.

\bibitem{[Sto06b]}
{G. Stoitcheva, W. Satu{\l}a, W. Nazarewicz, D.J. Dean, M. Zalewski and H.
  Zdu\'nczuk, Phys. Rev. {\bf C73}, 061304 (2006)}.

\bibitem{[Zal07aw]}
{M. Zalewski, W. Satu{\l}a, W. Nazarewicz, G. Stoitcheva, and H. Zdu\'nczuk,
  Phys. Rev. {\bf C75}, 054306 (2007)}.

\bibitem{[Sat08]}
{W. Satu{\l}a, R.A. Wyss and M. Zalewski, Phys. Rev. {\bf 78}, 011302(R)
  (2008)}.

\bibitem{[Sat07w]}
{W. Satu{\l}a, Int. J. Mod. Phys. {\bf E16}, 360 (2007)}.

\bibitem{[Les07]}
{T. Lesinski, M. Bender, K. Bennaceur, T. Duguet, and J. Meyer, Phys. Rev. C
  {\bf 76}, 014312 (2007)}.

\bibitem{[Toi08]}
{J. Toivanen, J. Dobaczewski, M. Kortelainen, and K. Mizuyama, Phys. Rev. C
  {\bf 78}, 034306 (2008)}.

\bibitem{[Klu09]}
{P. Kl\"upfel, P.-G. Reinhard, T.J. Burvenich, and J.A. Maruhn, Phys. Rev. {\bf
  C79}, 034310 (2009).}

\bibitem{[Ben09aa]}
{M. Bender, K. Bennaceur, T. Duguet, P.-H. Heenen, T. Lesinski, and J. Meyer,
  Phys. Rev. C, in press; arXiv:0909.3782}.

\bibitem{[Kor08]}
{M. Kortelainen, J. Dobaczewski, K. Mizuyama, and J. Toivanen, Phys. Rev. C
  {\bf 77}, 064307 (2008)}.

\bibitem{[Sat08a]}
{W. Satu{\l}a, J. Dobaczewski, W. Nazarewicz, and M. Rafalski, to be
  published}.

\end{thebibliography}

\end{document}